\documentclass{elsart}
\journal{Physics Letters B}
\usepackage{cite}
\usepackage{amssymb}
\usepackage{subeqn}
\usepackage{psfig}
\def \be{\begin{equation}}
\def \ee{\end{equation}}
\def \bea{\begin{eqnarray}}
\def \eea{\end{eqnarray}}
\def \ben{\begin{enumerate}}
\def \een{\end{enumerate}}
\def \bq{B_q}
\def \bd{B_d}
\def \bs{B_s}
\def \bem#1{\renewcommand{\thefootnote}{\arabic{footnote}}\footnote{#1}}

\def \braket#1#2#3{\langle #1|#2| #3\rangle}

\def \cl#1{#1\%\ \mathrm{C.L.}}

\def \cp{\mathrm{CP}}

\def \diag{\mathrm{diag}}

\def \ea{\emph{et al.}}
\def \eq#1{Eq.~(\ref{#1})}
\def \eqs#1#2{Eqs.~(\ref{#1})--(\ref{#2})}
\def \fig#1{Fig.~\ref{#1}}

\def \GeV{{\mathrm{GeV}}}

\def \hc{\mathrm{H.c.}}
\def \heff{H_{\mathrm{eff}}}
\def \heffhdm{\heff^{\mathrm{New}}}
\def \heffsm{\heff^{\mathrm{SM}}}
\def \Im{{\mathrm{Im}}\,}

\def \L{{\mathcal L}}

\def \Mhdm{M_{12}^{\mathrm{New}}}
\def \Msm{M_{12}^{\mathrm{SM}}}
\def \MeV{{\mathrm{MeV}}}
\def \nnu{\nonumber}

\def \O{{\mathcal O}}
\def \ol#1{\overline{#1}}
\def \ps{\mathrm{ps}}

\def \Re{{\mathrm{Re}}\,}
\def \rf{Ref.~\cite}
\def \rfs{Refs.~\cite}
\def \sec#1{Sec.~\ref{#1}}

\def \TeV{{\mathrm{TeV}}}
\def \Tr{{\mathrm{Tr}}\,}

\def \vckm{V_{\mathrm{CKM}}}
\newcommand{\TO}[2]{\stackrel {#1}{\hbox to #2pt{\rightarrowfill}}}
\def \a{\alpha}
\def \b{\beta}
\def \D{\Delta}
\def \g{\gamma}
\def \G{\Gamma}
\def \d{\delta}
\def \epsi{\epsilon}

\def \l{\lambda}

\def \m{\mu}

\def \p{\pi}
\def \r{\rho}
\def \s{\sigma}

\def\annu#1#2#3{{\it Annu.~Rev.~Nucl.~Part.~Sci.\/} {\bf #1} (19#2) #3}

\def\euro#1#2#3{{\it Eur.~Phys.~J.\/} {\bf C#1} (19#2) #3}
\def\ijmp#1#2#3{{\it Int.~J.~Mod.~Phys.\/}~{\bf A#1} (19#2) #3}

\def\np#1#2#3{{\it Nucl.~Phys.\/}~{\bf B#1} (19#2) #3}

\def\pl#1#2#3{{\it Phys.~Lett.\/}~{\bf B#1} (19#2) #3}
\def\plold#1#2#3{{\it Phys.~Lett.\/}~{\bf #1B} (19#2) #3}
\def\pp{Report No.~}
\def\prd#1#2#3{{\it Phys.~Rev.\/}~{\bf D#1} (19#2) #3}
\def\prl#1#2#3{{\it Phys.~Rev.~Lett.\/}~{\bf #1} (19#2) #3}

\def\ptp#1#2#3{{\it Prog.~Theor.~Phys.\/}~{\bf #1} (19#2) #3}

\def\rmp#1#2#3{{\it Rev.~Mod.~Phys.\/}~{\bf #1} (19#2) #3}
\def\zpc#1#2#3{{\it Z.~Phys.\/}~{\bf C#1} (19#2) #3}
\begin{document}
\begin{frontmatter}
\begin{flushright}
FISIST/4-99/CFIF\\hep-ph/9904379 \\ April 1999   
\end{flushright}
\title{A Real CKM Matrix and Physics Beyond the Standard Model}  
\date{}
\author{G.\,C. Branco\thanksref{gustavo}},
\author{F. Cagarrinho\thanksref{fernando}} 
and 
\author{F. Kr\"uger\thanksref{frank}}
\address{Centro de F\'\i sica das Interac\c{c}\~{o}es Fundamentais (CFIF),
Departamento de F\'{\i}sica,  Instituto Superior T\'ecnico,  
Av. Rovisco Pais,  1049-001 Lisboa, Portugal}
\thanks[gustavo]{E-mail address: d2003@beta.ist.utl.pt}
\thanks[fernando]{E-mail address: fernando@gtae2.ist.utl.pt}
\thanks[frank]{E-mail address: krueger@gtae3.ist.utl.pt}
\begin{abstract}
We study the possible existence of a real 
Ca\-bib\-bo-Ko\-ba\-ya\-shi-Mas\-ka\-wa 
(CKM) matrix, with $\cp$ violation originating from physics beyond the 
standard model (SM). We show that present experimental data allow for a 
real CKM matrix provided that new physics also contributes 
to $\D m_{B_d}$ by at least $20\%$ of the SM contribution  (for $\r > 0$), 
besides generating $\cp$ violation in the kaon sector. 
The naturalness of a real CKM matrix is studied
within  the framework of general multi-Higgs-doublet models with 
spontaneous $\cp$ violation. 
As an example, we discuss a specific two-Higgs-doublet model and its 
implications for $\cp$ asymmetries in non-leptonic neutral $B$-meson decays.
\end{abstract}
\end{frontmatter}
\setcounter{footnote}{0} 
%
%%%%%%%%%%%%%% END OF TITLE PAGE %%%%%%%%%%%%%%%%%%%%%%
%
\section{Introduction}
At present all information on $\cp$ violation is consistent with the 
standard model (SM) of electroweak interactions 
and its Cabibbo-Kobayashi-Maskawa (CKM) mechanism 
of $\cp$ breaking \cite{ckm}. 

In this Letter, we address the question of    
whether or not current experimental data,
in particular the new limit for the $B_s^0$ oscillation frequency, 
exclude the possibility of having a real CKM matrix. 
We envisage a scenario where charged weak interactions are $\cp$ conserving, while all $\cp$-violating phenomena stem from physics beyond the SM. 
Along with unitarity, the CKM matrix 
$\vckm$ is already constrained by experimental data; in particular 
strange particle and $B$-meson decays lead to the measurement of 
the moduli of the CKM matrix elements 
$|V_{us}|$, $|V_{cb}|$, and $|V_{ub}/V_{cb}|$.
Since the new physics contribution to the above decays in the most 
plausible extensions of the standard theory cannot compete 
with the SM tree-level processes, the determination of these matrix elements 
is likely to hold, even in the pre\-sence of physics beyond the SM 
\cite{nirquinn}.
Further constraints on the CKM matrix arise from the strength of $\cp$ violation in the kaon sector measured by $|\epsi_K|$ and from the observation 
of $B_d^0$--$\bar{B}_d^0$ mixing, along with the improved experimental
lower limit of 
$\D m_{B_s}$. Within the SM, the experimental results for 
$\D m_{B_d}$ and $\D m_{B_s}$ can be used to extract the values of $|V_{td}|$ and $|V_{ts}|$, while the measured value of $|\epsi_K|$ 
directly constrains the size of the CKM phase.

What follows is a simple method of checking whether or not present 
experimental data already 
exclude the possibility of having a real CKM matrix. 
Let us assume that $\vckm$ is real\bem{This case has also been discussed, 
for example, in \rf{relpap}.} 
and that the observed 
$\cp$ violation in the kaon sector arises solely from physics beyond the SM. 
The experimental information on  $|V_{us}|$, $|V_{cb}|$, and 
$|V_{ub}/V_{cb}|$, together with the unitarity of $\vckm$, allows one to 
deduce the values for $|V_{td}|$ and $|V_{ts}|$. 
Moreover, a comparison of these values
with those extracted from $\D m_{B_d}$ and $\D m_{B_s}$ measurements   
enables us to ascertain the viability of a real CKM matrix. One may be 
tempted to interpret any incompatibility between the two sets of $|V_{td}|$ 
and $|V_{ts}|$ values obtained above as an indication that the CKM matrix 
cannot be real. However, the situation is more involved 
for the following reason. 
Within the SM, $\D m_{B_d}$ and $\D m_{B_s}$ are generated only at one-loop level by the 
$W$-mediated box diagrams, and as a result, physics beyond the SM can 
significantly contribute to $\D m_{B_d}$ and  $\D m_{B_s}$. Indeed, 
in many extensions of the standard theory, 
the same new physics that gives rise to
$\epsi_K$ is likely to provide significant contributions 
to $\D m_{B_d}$ and $\D m_{B_s}$.

In this paper, we first analyse the somewhat unnatural situation where new physics contributes only to $\epsi_K$ and not to the mass 
difference in the $B$ system. We show that in this case the currently 
available data already disfavour a real CKM matrix, thus confirming previous results 
\cite{che:etal,parodi:etal}. 
We then investigate the more plausible situation of a real CKM matrix, with new physics generating $\epsi_K$ and also contributing to    
$\D m_{B_d}$ and $\D m_{B_s}$. In this case, as we shall see below, 
present experimental data are compatible with the scenario of 
a real CKM matrix. 

Our paper is organized as follows.~In \sec{general}, we present a 
general analysis of $\D B=2$ transitions in the presence of new physics. 
Our results are model independent and apply to a wide class of models 
with a real CKM matrix. 
Section \ref{general:realCKM} deals with multi-Higgs-doublet 
models in which $\cp$ is softly broken. Assuming that $\cp$ is a symmetry 
of the Lagrangian, spontaneously broken by the vacuum, 
we discuss the possibility of having a real CKM matrix in a natural way.
In \sec{2HDM}, we illustrate how the 
experimental constraints on $\epsi_K$, 
$\D m_K$, and $\D m_{B_d}$ can be satisfied with a real CKM matrix
in the context of a simple model with two Higgs doublets, and 
analyse the predictions for $\cp$ asymmetries in non-leptonic 
neutral $B$ decays that will be measured in upcoming $B$ experiments.
Finally, we present our conclusions in \sec{summary}.

\section{General analysis}\label{general}
\subsection{Basic formulae and notation}
We first carry out a general analysis of the new physics 
contribution to $\D B=2$ transitions (see, e.g., 
\rfs{babar,grossmanetal}).
The off-diagonal element of the $B_q$ mass matrix can be written as\bem{For notational simplicity, we will  omit the $q$ 
subscript for $\heff$.} 
\be\label{mdef}
M_{12}(B_q)\equiv \Mhdm(B_q) + \Msm(B_q)
=\frac{\braket{B_q}{\heff}{\bar{B}_q}}{2m_{B_q}}\ ,
\ee
where $q=d$ or $s$, and the effective Hamiltonian has the form
\be\label{fullheff}
\heff = \heffsm + \heffhdm\ .
\ee 
It is customary to parametrize the new physics contribution appearing in 
\eq{mdef} through
\be\label{np}
r_q^2 \exp(2i\theta_q) \equiv
\frac{M_{12}(B_q)}{\Msm(B_q)}\ ,
\ee
so that
\be
\D m_{\bq}= 2|\Msm|_q r^2_q\ .
\ee
Note that $r_q^2< 1$ arises if the new physics amplitudes  interfere 
destructively with those of the SM. The contribution of the SM
to $B_q^0$--$\bar{B}_q^0$ mixing is given by
\be\label{m12sm}
\Msm (\bq)=\frac{G_F^2}{12\p^2}\eta_B m_{\bq}m_W^2 f_{B_q}^2B_{B_q}S_0(x_t)
(V_{tq}^{}V_{tb}^*)^2, \quad x_t = \ol{m}_t^2/m_W^2\ ,   
\ee
where the various factors appearing in \eq{m12sm} 
may be found elsewhere \cite{qcd:factors,buras}. In fact, we have
\be\label{m12:bd}
\Msm (\bd)= 3.1\times 10^3\ \ps^{-1}
\left[\frac{f_{B_d}\sqrt{B_{B_d}}}{200\ \MeV}\right]^2
\left[\frac{S_0(x_t)}{2.36}\right]
\left[\frac{\eta_B}{0.55}\right]
(V_{td}^{}V_{tb}^*)^2\ ,
\ee
and 
\be
\Msm (\bs) = 4.5\times 10^3\ \ps^{-1}
\left[\frac{f_{B_s}\sqrt{B_{B_s}}}{240\ \MeV}\right]^2
\left[\frac{S_0(x_t)}{2.36}\right]
\left[\frac{\eta_B}{0.55}\right] 
(V_{ts}^{}V_{tb}^*)^2\ .
\ee
Introducing the SU(3) breaking parameter
\be
\xi_s\equiv \frac{f_{B_s}\sqrt{B_{B_s}}}{f_{B_d}
\sqrt{B_{B_d}}}\ ,
\ee
we find the ratio
\be\label{rdrs}
\frac{\D m_{B_s}}{\D m_{B_d}}
=\xi^2_s\frac{m_{B_s}}{m_{B_d}} 
\left|\frac{V_{ts}}{V_{td}}\right|^2\frac{r_s^2}{r_d^2}\ .
\ee
Finally, the  $\cp$-violating asymmetry between $B^0$ and $\bar{B}^0$ 
mesons decaying to $\cp$ ei\-gen\-states $f_\cp$, 
such as $J/\psi K^0_S$ or $\p^+\p^-$,
in the presence of new physics is given by
\be\label{def:asym}
a_{f_\cp}(t)=\frac{\G(B^0_d(t)\to f_\cp)-\G(\bar{B}^0_d(t)\to f_\cp)}{\G(B^0_d(t)\to f_\cp)+\G(\bar{B}^0_d(t)\to f_\cp)}= -a_{f_\cp} \sin (\D m_{B_d}t)\ ,
\ee
with
\be\label{asym}
a_{J/\psi K^0_S}= \sin 2(\b +\theta_d+ \theta_K), \quad 
a_{\p^+\p^-} = \sin 2(\a -\theta_d)\ .
\ee
Here we have ignored penguin contributions as well as 
final-state interaction phases.
In addition, we have assumed that the direct decay 
amplitudes are dominated by the SM tree diagrams. 
The phase $\theta_K$ in the above expression is caused by 
the new physics contribution 
(normalized to the SM amplitude) to $K^0$--$\bar{K}^0$ mixing, i.e. the mixing parameter 
$(q/p)_K$ picks up another phase. However, given the experimental information 
on $\epsi_K$, we expect the phase $\theta_K$ to be of $\O(10^{-3})$. 
We will return to this point in \sec{2HDM}. 

\subsection{New physics contribution and experimental data}
\begin{figure}
\centerline{\psfig{figure=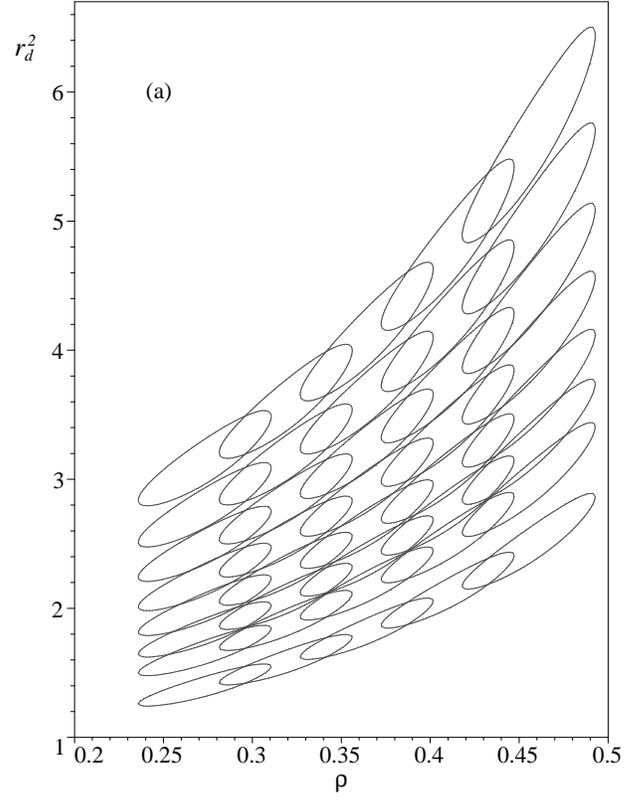,width=3.2in,angle=0}}
\vspace{0.4cm}
\centerline{\psfig{figure=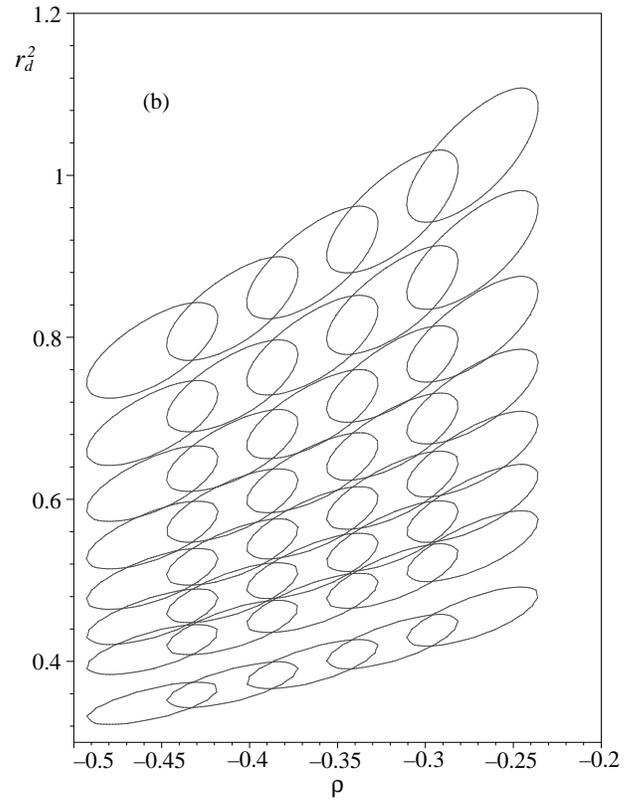,width=3.2in,angle=0}}
\caption[]{The allowed region (at $\cl{95}$) in the $(\r, r_d^2)$ plane for 
the Wolfenstein parameter $\r>0$ (a) and $\r<0$ (b), using 
the experimental values for $|V_{ub}/V_{cb}|$, $|V_{cb}|$,  and $\D m_{B_d}$. 
Each contour corresponds to a variation of the theoretical input 
parameters, as described in the text [cf.~\eq{theo:var1}].\label{fig1}}
\end{figure}

As mentioned above, we will deal with a real CKM matrix. 
In this case, we have 
\be
\left|\frac{V_{ub}}{V_{cb}}\right|= \l |\r|\ ,
\ee
and (neglecting higher order terms in $\l$) 
\be
V_{td}\simeq A\l^3\left[1-\r\left(1-\frac{1}{2}\l^2\right)\right],\quad 
V_{ts}\simeq -A\l^2\left[1-\frac{1}{2}\l^2(1-2\r)\right]\ ,
\ee
with the Wolfenstein parameters $A$, $\l$, and $\r$ \cite{ckm:wolfenstein}. 
In our numerical analysis, we use the  following input 
parameters \cite{babar,buras,pdg,alexander,sachrajda}:
\be
\left|\frac{V_{ub}}{V_{cb}}\right| =
\left|\frac{V_{ub}}{V_{cb}}\right|_{\mathrm{T}}\pm 0.005,
\quad |V_{cb}| = 0.0395 \pm 0.0017\ ,
\ee
\be
A= 0.819 \pm 0.035, \quad \l = 0.2196\pm 0.0023, 
\quad \ol{m}_t= 167\pm 6\ \GeV\ ,
\ee
\be\label{exp:epsi}
|\epsi_K| = (2.280\pm 0.019)\times 10^{-3}\ ,
\ee
\be\label{exp:deltamk}
\D m_K= (0.5301\pm 0.0014)\times 10^{-2}\ \ps^{-1}\ , 
\ee
\be\label{bound:bs}
\D m_{B_d}= 0.471\pm 0.016\ \ps^{-1},\quad 
\D m_{B_s}> 12.4\ \ps^{-1}\ (\cl{95})\ ,
\ee
and consider the ranges 
\be\label{theo:var1}
0.06\leqslant \left|\frac{V_{ub}}{V_{cb}}\right|_{\mathrm{T}}
\leqslant 0.10,\quad
160\ \MeV\leqslant f_{B_d}\sqrt{B_d}\leqslant 240\ \MeV\ ,
\ee
\be\label{theo:var2}
1.12\leqslant \xi^2_s\leqslant 1.48\ .
\ee
By using the experimentally measured values for $|V_{ub}/V_{cb}|$, $|V_{cb}|$,  and $\D m_{B_d}$ (ignoring for the moment the 
constraint imposed by the $B_s$ system), we can perform a fit 
with two parameters $\r$ and $r_d^2$. The results of the fit are shown in 
\fig{fig1}, 
%
%%%%%%%%%%%% Figure 1 %%%%%%%%%%%%%%%%%%%%%%%%%%%%%%%%%%%%%%%%%%%%%%%%%%%%
%
%
%%%%%%%%%%%%%%%%%%%%%%%%%%%%%%%%%%%%%%%%%%%%%%%%%%%%%%%%%%%%%%%%%%%%%%%%%%
%
from which we infer the upper and lower limits 
\begin{subequations}\label{bound:rho}
\be
0.24\lesssim \r \lesssim 0.49, \quad 1.2 \lesssim r_d^2 \lesssim 6.5\ ,
\ee
\be
-0.49\lesssim \r \lesssim -0.24, \quad 0.3\lesssim r_d^2 \lesssim 1.1\ .
\ee
\end{subequations}
Thus, for positive $\r$ values, a new physics contribution 
to  $\D m_{B_d}$ of at least $20\%$ of the SM contribution is required for 
consistency with current experimental data.
Including the $\D m_{B_s}$ constraint, our analysis gives the lower bound as 
\be\label{lower}
\r > 1-\sqrt{C_s\frac{r_s^2}{r_d^2}},\quad 
C_s = \frac{1}{\l^2}\frac{m_{B_s}}{m_{B_d}}
\frac{\D m_{B_d}}{\D m_{B_s}}\,\xi_s^2\leqslant 0.8\, \xi_s^2\ .
\ee
Two remarks are in order here. First, it is 
important to note  that without new physics contributions to $\D m_{B_d}$ and 
$\D m_{B_s}$, 
i.e.~$r_d^2= r_s^2=1$, we obtain from \eq{lower} $\r> - 0.08$.
By using the bound of \eq{bound:rho}, one immediately sees that a real 
CKM matrix is disfavoured by present experimental data, as also
reported in \rfs{che:etal,parodi:etal}.
Second, for $r_s^2=1$ but $r_d^2\neq 1$, it follows from  
Eqs.~(\ref{bound:rho}) and (\ref{lower}) that  
either negative or positive values for $\r$ are allowed by current
experimental data. That is, a real CKM matrix is still consistent 
with the above measurements provided new physics contributes to both the $K$
and $B_d^0$ system. 
 
\section{A real CKM matrix}\label{general:realCKM}
\subsection{Multi-Higgs-doublet models}
This section is concerned with the question of how to construct models that 
naturally lead to a real CKM matrix, within  the framework of 
multi-Higgs-doublet models (MHDMs).\bem{For a concise review of models with 
$\cp$ violation arising from the Higgs sector see, e.g., \rf{bigi}.}

The most general Yukawa couplings in the MHDM are given by the following
Lagrangian (in the weak eigenbasis)
\be\label{yukawa:weak}
\L_Y=-\sum_{j=1}^{M}\left(\bar{Q}_L\G_j \phi_j n_R +\bar{Q}_L\tilde{\G}_j 
\tilde{\phi}_j p_R\right) +\hc\ , 
\ee 
with the left-handed quark doublet $Q_L = (p_L, n_L)^T$ 
and the right-handed quark singlets $p_R$ and $n_R$,
where $p$ and $n$ denote the up- and down-type quarks respectively. 
$\G_j$ and $\tilde{\G}_j$ are matrices in flavour space, and the 
Higgs doublets are given by $\phi_j = (\phi_j^+, \phi_j^0)^T$, 
$\tilde{\phi}_j = i\s_2 \phi^*_j$, with $j=1, \dots, M$.
In order to obtain a real CKM matrix in a natural way,  
it is particularly important to require $\cp$ invariance of the Lagrangian,
only spontaneously broken by the vacuum. Without loss of generality,
we may therefore assume that the above Yukawa matrices are real, 
while the vacuum expectation values (VEVs) of the neutral Higgs fields are given by
$\langle{\phi_j^0}\rangle = v_j \exp(i\a_j)$, with 
$v^2=\sum_j v_j^2= (\sqrt{2}G_F)^{-1}$. Defining the Hermitian matrices
\bea\label{hu}
H_u &\equiv& M_u M_u^{\dagger}
= \frac{1}{2}\bigg\{\sum_i v_i^2\tilde{\G}_i\tilde{\G}_i^T 
+ \sum_{j>i}v_i v_j
\Big[\cos(\a_j-\a_i)(\tilde{\G}_i\tilde{\G}_j^T+
\tilde{\G}_j\tilde{\G}_i^T)\nnu\\
&&\mbox{}\hspace{2.15cm}+i \sin(\a_j-\a_i)
(\tilde{\G}_i\tilde{\G}_j^T-\tilde{\G}_j\tilde{\G}_i^T)\Big]\bigg\}\ ,
\eea
and
\bea\label{hd}
H_d &\equiv& M_d M_d^{\dagger}
= \frac{1}{2}
\bigg\{\sum_i v_i^2\G_i\G_i^T 
+ \sum_{j>i}v_i v_j \Big[\cos(\a_j-\a_i)(\G_i\G_j^T+\G_j\G_i^T)\nnu\\
&&\mbox{}\hspace{2.15cm}+i \sin(\a_j-\a_i)(\G_j\G_i^T-
\G_i\G_j^T)\Big]\bigg\}\ ,
\eea
where $M_u$ and $M_d$ are  the non-diagonal up- and down-type
mass matrices respectively, one can show that within the three-generation SM 
a necessary and sufficient condition for 
$\cp$ invariance in the charged gauge interactions
is given by \cite{commutator}
\be\label{condition}
T\equiv \Tr[H_u, H_d]^3=0\ .
\ee
Note that this is independent of the number of Higgs doublets which generate 
masses for the fer\-mi\-ons. Hence, $T\neq 0$ implies $\cp$ violation 
mediated by charged gauge interactions, 
while $T=0$ guarantees a real CKM matrix. 

From \eq{condition} it is evident that the simplest and most natural way 
of obtaining a real CKM matrix (i.e. without any fine-tuning)
is by introducing additional symmetries with the effect that
\be\label{relation:gamma}
\G_j\G_i^T=\G_i\G_j^T 
\quad \mathrm{and}\quad \tilde{\G}_i\tilde{\G}_j^T
=\tilde{\G}_j\tilde{\G}_i^T\ .
\ee
Various multi-Higgs-doublet models where the condition in 
\eq{relation:gamma} is satisfied as a result of a discrete 
symmetry have been discussed in the literature. 
For example, the simplest models belonging to this class contain two Higgs 
doublets and a $Z_2$ symmetry \cite{luis} or three Higgs doublets and a 
$Z_3$ symmetry \cite{gustavo-sanda}. In these models, there are 
flavour-changing neutral currents and the vacuum leads to spontaneous $\cp$ 
violation, but the CKM matrix is real as a result of the discrete symmetry.

\subsection{Supersymmetric standard model}
The supersymmetric SM  with spontaneous $\cp$ violation (SCPV) is another 
example where the CKM matrix is real provided that only two Higgs doublets are 
introduced. In the minimal supersymmetric standard model (MSSM), the tree-level
vacua are all $\cp$ conserving so that SCPV can only occur if 
radiative corrections to the Higgs potential are taken into account. 
This inevitably leads to a mass of the lightest neutral Higgs boson 
which has already been ruled out by experiment \cite{susy:mssm}.
On the other hand, in the next-to-minimal supersymmetric standard model 
(NMSSM) with one or more singlet fields in addition to the two doublets 
\cite{susy:nmssm}, SCPV can occur even at tree level. 
Whether or not this scenario is still consistent with present 
experimental bounds 
on the Higgs mass depends on the specific model, and we refer 
the interested reader to \rf{disc:nmssm}. 
The important point to emphasize is that in the supersymmetric SM, 
provided there are only two Higgs doublets, but allowing for an arbitrary 
number of singlets, the CKM matrix is real, even if there is a physically 
meaningful relative phase between the  VEVs of the two doublets. 
This can be readily verified  by noting that the above phase can be eliminated 
from the quark mixing matrix \cite{gustavo}.

\section{A specific model}\label{2HDM}
\subsection{Two-Higgs-doublet model}
For definiteness, we will consider a minimal extension of the SM
with SCPV and a real CKM matrix as a result of a $Z_2$ symmetry \cite{luis}.
Our main aim is to show that, within the framework of a specific 
two-Higgs-doublet model (2HDM),
it is  possible to have a real CKM matrix in a natural way, with 
$\epsi_K$ and $\D m_{B_d}$ generated by physics beyond the SM, while
taking into account the constraints discussed in \sec{general}. 
As we shall see below, $\cp$ violation arises solely 
from the exchange of neutral Higgs bosons, 
thereby inducing flavour-changing neutral current (FCNC) interactions at 
tree level.

We start  with the mass matrices $M_u$ and $M_d$ 
which can be diagonalized by means of a biunitary transformation, namely
\be
p_L\to V_L^u u_L, \quad n_L\to V_L^d d_L, \quad p_R\to V_R^u u_R, 
\quad n_R\to V_R^d d_R\ ,
\ee 
so that
\be\label{massmatrix}
M_d^{\diag}= V_L^{d\dagger}M_d V_R^d, \quad 
M_u^{\diag}= V_L^{u\dagger}M_u V_R^u\ ,
\ee
where 
\be
M_u= \frac{1}{\sqrt{2}}(v_1 \tilde{\G}_1 +v_2 e^{-i\a} \tilde{\G}_2), \quad
M_d= \frac{1}{\sqrt{2}}(v_1 \G_1 +v_2 e^{i\a}\G_2)\ .
\ee

In order to  obtain the physical Higgs fields, 
it is useful to parametrize $\phi_1$ 
and $\phi_2$ in \eq{yukawa:weak} as follows
\be
\phi_1=\frac{1}{\sqrt{2}} \left(\begin{array}{c} \sqrt{2}\varphi_1^+ \\ 
v_1+ \r_1+i \eta_1
\end{array}\right) , \quad 
\phi_2=\frac{e^{i\a}}{\sqrt{2}} \left(\begin{array}{c} \sqrt{2}\varphi_2^+ \\ 
v_2+ \r_2+i \eta_2\end{array}\right) .
\ee
The pseudo-Goldstone bosons $G^+$ and $G^0$ in our model can then be found 
by introducing new bases, i.e.
\be
\left(\begin{array}{c}G^+\\H^+\end{array}\right)= O
\left(\begin{array}{c}\varphi_1^+\\ \varphi_2^+\end{array}\right) , \quad
\left(\begin{array}{c}G^0\\ I\end{array}\right)= O
\left(\begin{array}{c}\eta_1\\ \eta_2\end{array}\right) , \quad
\left(\begin{array}{c}H^0\\ R\end{array}\right)= O
\left(\begin{array}{c}\r_1\\ \r_2\end{array}\right) ,
\ee
with
\be
O= \frac{1}{v}\left(\begin{array}{lr}v_1 & v_2\\ v_2& -v_1\end{array}\right)\ .
\ee
The mass matrix of the neutral Higgs fields is 
diagonalized through an orthogonal
matrix $U$ relating $H^0$, $R$, and $I$ to the mass eigenstates $H_i$
($i=1, 2, 3$) via the 
relations
\be
H^0 = \sum_{i=1}^3 U_{1i}H_i, \quad R = \sum_{i=1}^3 U_{2i}H_i, \quad
I = \sum_{i=1}^3 U_{3i}H_i\ .
\ee

The Higgs-boson interactions with the quarks are then governed 
by the following Lagrangian
\bea\label{yukawa:mass}
\L_Y&=& (2\sqrt{2}G_F)^{1/2}
\Bigg\{\Big[\bar{u}\Big(\G^{u\dagger}
\vckm P_L-\vckm\G^d P_R\Big)d\Big] H^+ + \hc\Bigg\}\nnu\\
&+&(\sqrt{2}G_F)^{1/2}\sum_{i=1}^3
\Bigg\{-U_{1i}H_i\Big[\bar{d}M_d^{\diag}d+\bar{u}M_u^{\diag}u\Big]\nnu\\
&&\hspace{1cm}\mbox{}-U_{2i}H_i\Big[\bar{d}\big(\G^d P_R+ \G^{d\dagger} P_L\big)d
+ \bar{u}\big(\G^u P_R + \G^{u\dagger} P_L\big)u\Big]\nnu\\
&&\hspace{1cm}\mbox{}+ iU_{3i}H_i\Big[\bar{u}\big(\G^u P_R - \G^{u\dagger} P_L\big)u
- \bar{d}\big(\G^d P_R - \G^{d\dagger} P_L\big)d\Big]\Bigg\}\ ,
\eea
where $\vckm\equiv V_L^{u\dagger}V_L^d$ is the usual CKM matrix, 
$P_{L,R}=(1\mp\g_5)/2$, and
\bea
\G^u=V_L^{u\dagger}\left(\tilde{\G}_1\frac{v_2}{\sqrt{2}}
-e^{-i\a}\tilde{\G}_2\frac{v_1}{\sqrt{2}}\right)V_R^u,\ 
\G^d=V_L^{d\dagger}\left(\G_1\frac{v_2}{\sqrt{2}}
-e^{i\a}\G_2\frac{v_1}{\sqrt{2}}\right)V_R^d\ .\nnu\\
\eea
In order for the above Yukawa coupling matrices to
fulfil the relation of \eq{relation:gamma}, we impose an extra discrete $Z_2$ 
symmetry on the  quark sector [see \eq{yukawa:weak}] under which  
$\phi_2\to -\phi_2$ and $n_{R3}\to -n_{R3}$ ,
while all other fields remain unchanged.\bem{An alternative assignment 
is to choose both $n_{R3}$ as well as $p_{R3}$ to be odd under $Z_2$. 
The advantage of this choice is that it leads to a naturally  
vanishing parameter $\bar{\theta}\equiv \theta_{\mathrm{QFD}} +
\theta_{\mathrm{QCD}}$ at tree level, 
where $\theta_{\mathrm{QFD}}= \arg[\det(M_u M_d)]$.}   
Denoting the ratio of the VEVs of the two Higgs bosons 
by $\tan\b\equiv v_2/v_1$,  we find
\be\label{gammau:diag}
\G^u =  M_u^{\diag}\tan\b\ .
\ee
Hence there are no FCNC interactions in the up-quark sector
(i.e. no $\cp$ violation in $D^0$--$\bar{D}^0$ mixing).

Turning to the down-quark sector, the CKM matrix can be made 
real\bem{The one-loop corrections to the parameter $J_{\cp}$ 
\cite{jarlskog} are
equal to zero within the 2HDM in question \cite{luis:private}.} 
by redefining the phase of the quark field through 
$n_{R3}\to e^{-i \a}n_{R3}$ 
so that $V_L^{u,d}$ and $V_R^{u,d}$ are real and orthogonal matrices, 
while $\G_d$ takes the explicit form
\be\label{gammad}
\G^d= \left(\begin{array}{ccc} m_d(\tan\b-Sx^2) & -m_d S\b_K & -m_d S\b_d\\
-m_s S\b_K & m_s(\tan\b-Sy^2) & -m_s S\b_s\\
-m_b S\b_d & -m_b S\b_s & m_b(\tan\b-Sz^2) 
\end{array}\right) .
\ee
Here $S=\tan\b + \cot\b$, with $S \geqslant 2$, and the 
$\b_n$ $(n=K, d, s)$ are defined as
\be\label{beta:def}
\b_K= xy, \quad \b_d= xz, \quad \b_s=yz\ ,
\ee
where $x, y, z$ refer to the elements of the third column of 
$V_R^{d\dagger}$ with $x^2+y^2+z^2=1$.
Consequently, 
the flavour-changing couplings of the three real neutral Higgs fields 
are constrained to obey
\be\label{beta:constraint}
\b_K\b_d\b_s = \left(\b_K\b_d\right)^2+\left(\b_K\b_s\right)^2
+\left(\b_d\b_s\right)^2, \quad |\b_n|\leqslant 1/2\ .
\ee
The crucial point here is that the mass matrix $M_d$, given in \eq{massmatrix},
is in general not a Hermitian matrix and consequently 
$V_R^d\neq V_L^d$,
which in turn implies that $V_R^d$ is unrelated to $\vckm$. As a result, the 
$\b_n$ above can be taken as free parameters, constrained only by 
condition (\ref{beta:constraint}). 
This is an important feature because the 
strength of the neutral Higgs contributions to $\D m_K$ and $\D m_{B_{d, s}}$
(see below) are proportional to $\b_K$ and $\b_{d, s}$ 
respectively.
In fact, the neutral Higgs exchange induces a new physics contribution,
in addition to the SM effective $\D S = 2$ Hamiltonian in \eq{fullheff},  
given by
\bea\label{szwei}
\lefteqn{\heffhdm =-\frac{G_F}{\sqrt{2}}\,m_s^2 (\b_K S)^2 
\sum_{i = 1}^{3}\frac{1}{M_i^2}}&&\nnu\\
&&\times\bigg\{ C_i^{LL}(\bar{d}P_L s)^2 + C_i^{RR}(\bar{d}P_Rs)^2 
+C_i^{LR}(\bar{d}P_Ls)(\bar{d}P_R s)\bigg\} + \hc\ ,
\eea
with
\bea
C_i^{LL}= 2(U_{2i}-iU_{3i})^2,\ C_i^{RR}= 2\zeta^2(U_{2i}+iU_{3i})^2,
\  C_i^{LR}= 4 \zeta (U_{2i}^2+U_{3i}^2)\ ,\nnu\\
\eea
where we have introduced the shorthand $\zeta = m_d/m_s$. 
The effective Ha\-mil\-to\-nian inducing $B_q^0$--$\bar{B}_q^0$ mixing
can be obtained  from the above by replacing 
\bea
\label{replacement}
s\to b,\ m_s \to m_b,\ \b_K\to \b_d;\quad
s\to b,\ m_s \to m_b,\ m_d \to m_s,\ \b_K\to \b_s\ ,\nnu\\
\eea
for the $B_d^0$ and $B_s^0$ system respectively.

To estimate the order of magnitude of the hadronic matrix elements resulting 
from \eq{szwei}, we rely on the 
vacuum insertion approximation. 
Defining the $F$-meson decay constant $f_F$ ($F=K, B_q$) via
\be
\braket{F^0}{\bar{q}\g_{\m}\g_5q'}{0}\braket{0}{\bar{q}\g^{\m}\g_5q'}{\bar{F}^0} = f_F^2 m_F^2, \quad q= d, s,\ q' = b, s,\ q'\neq q\ ,
\ee
and employing the equation of motion, we get
\be
\braket{F^0}{\bar{q}\g_5q'}{0}\braket{0}{\bar{q}\g_5q'}{\bar{F}^0} 
= - \left(\frac{m_F}{m_q + m_{q'}}\right)^2 f_F^2 m_F^2 \ .
\ee
Thus, we obtain for the relevant matrix elements 
of the four-quark operators  
\be
\braket{F^0}{(\bar{q}P_Lq')(\bar{q}P_Rq')}{\bar{F^0}}=
\frac{1}{2}\left[\left(\frac{m_F}{m_q + m_{q'}}\right)^2 + \frac{1}{6}\right]
f_F^2B_{1F} m_F^2\ , 
\ee
\be
\braket{F^0}{(\bar{q}P_{L,R}q')(\bar{q}P_{L,R}q')}{\bar{F^0}}=
-\frac{5}{12}\left(\frac{m_F}{m_q + m_{q'}}\right)^2 f_F^2 B_{2F}m_F^2\ ,
\ee
where $B_{iF}$ are the bag parameters. In what follows we will  use 
$B_{iF}=1$, as de\-ter\-mined by the vacuum insertion approximation, which is 
sufficient for our purposes.

The new physics contribution to $M_{12}$ in the $K$ 
system, \eq{szwei},  then takes the form
\bea
\lefteqn{\left.\Mhdm\right|_K = \frac{G_F}{\sqrt{2}}\,
m_s^2(\b_K S)^2f_K^2 m_K \sum_{i=1}^3
\frac{1}{M_i^2}}&&\nnu\\
&\times&
\Bigg\{\Bigg(\frac{5}{12}\Big[(U_{2i}-iU_{3i})^2+ \zeta^2(U_{2i}+iU_{3i})^2\Big]
-\zeta(U_{2i}^2+U_{3i}^2)\Bigg)
\Bigg(\frac{m_K}{m_s + m_d}\Bigg)^2\nnu\\
&-&\frac{1}{6}\zeta (U_{2i}^2+U_{3i}^2)\Bigg\}\ ,
\eea
which in the limit $\zeta \ll 1$ reduces to
\be\label{m12:Knew}
\left.\Mhdm\right|_K\simeq \frac{5}{12}\,\frac{G_F}{\sqrt{2}}f_K^2 m_K^3 
(\b_K S)^2\sum_{i=1}^3\frac{1}{M_i^2}(U_{2i}^2-U_{3i}^2- 2iU_{2i}U_{3i})\ .
\ee
As can be seen from Eqs.~(\ref{replacement}) and (\ref{m12:Knew}), 
the phase in the off-diagonal element $M_{12}$ is common to both 
the $K$ and $B$ system, and we may therefore write
\begin{subequations}\label{m12:gen}
\be
M_{12}^n = |\Mhdm|_n e^{i\d}+ \Msm|_n\equiv \Msm|_n 
r^2_n e^{2i\theta_n}, \quad n= K, d, s\ , 
\ee
with
\be
\theta_n = \frac{1}{2}\arctan\left(\frac{R_n\sin\d}{1+R_n\cos\d}\right),
\quad
r^2_n = \sqrt{1+ 2R_n\cos\d+R^2_n}\ ,
\ee
\end{subequations}
where $R_n\equiv  |\Mhdm|_n/\Msm|_n$.  

\subsection{Constraints from $\epsi_K$, $\D m_K$, and $\D m_{B_d}$}
We now proceed to discuss various constraints on the parameters of the 2HDM.
The observed $\cp$ violation in $K^0$--$\bar{K}^0$ mixing 
is described by (for $\epsi'\ll\epsi_K$)
\be\label{formula:epsi}
\epsi_K\simeq\frac{e^{i\p/4}}{\sqrt{2}}
\left(\frac{\Im M_{12}^K}{\D m_K}\right)\ ,\ee
whereas the mass difference in the $K$ system has the form
\be\label{massdiff:kaons}
\D m_K\simeq 2|\Re M_{12}^K|\ .
\ee

Using the experimental information on $\epsi_K$ and $\D m_K$,
Eqs.~(\ref{exp:epsi}) and (\ref{exp:deltamk}),  as well as the central value
$f_K= 160 \ \MeV$, we obtain
\be\label{cpconstraint}
(\b_K S)^2\sum_{i=1}^3 \frac{U_{2i}U_{3i}}{(M_i/\TeV)^2}\simeq 
5.2 \times 10^{-4}\ ,
\ee
and 
\be\label{mass:diffK}
(\b_K S)^2 \sum_{i=1}^3\frac{U_{2i}^2-U_{3i}^2}{(M_i/\TeV)^2}
\lesssim 0.16\ .
\ee
A few remarks are in order here. First of all, it can be readily shown 
that \cite{luis} 
\be\label{cotalpha}
\sum_{i=1}^3\frac{U_{2i}U_{3i}}{M_i^2}\propto \cot\a\ .
\ee
Second, an analysis of the Higgs potential shows that in the limit of an
exact $Z_2$ symmetry, the vacuum does not violate $\cp$ and the minimum of the potential is at
$\a=\p/2$, so that $\cot\a =0$. Allowing only a soft breaking of the 
$Z_2$ symmetry by a dimension-two term in the Higgs potential, 
spontaneous $\cp$ violation can be generated \cite{luis,gi}. 
This in turn implies that one 
may have $\cot\a \ll 1$ in a natural way (in the sense of 't Hooft 
\cite{thooft}), since the limit $\cot\a =0$ corresponds to the 
restoration of an exact $Z_2$ symmetry of the Lagrangian. This is the 
mechanism proposed by Branco and Rebelo \cite{gi} in order to naturally 
suppress the strength of $\cp$ violation. Therefore, the limit of 
\eq{cpconstraint}, which is due to the $\epsi_K$ measurement, does not 
necessarily imply very large values of the Higgs mass. On the other hand, 
we do not consider any accidental cancellation between the elements
$U_{2i}$ and $U_{3i}$ appearing in \eq{mass:diffK} 
because it would require additional fine-tuning of 
the parameters in the  Higgs potential. Furthermore, we should emphasize here
that there are also contributions from box diagrams with  
$W$-boson and non-SM charged Higgs particle exchange, 
as well as long-distance contributions, which affect only the \emph{real part} 
of $M_{12}^K$. 
Thus  the most stringent limit on $M_i$ comes from the experimentally 
measured value of  $\D m_K$, resulting in the upper bound of  
\eq{mass:diffK}.
So, taking  $(\b_K S)$ to be of order unity, the Higgs masses  have to 
satisfy $M_i \gtrsim \mathrm{few}\ \TeV$, whereas if we assume  
$(\b_K S) \sim \O(10^{-1})$, then it is sufficient to demand that 
$M_i\gtrsim  250\ \GeV$.

Turning to the $B$ system, the mass difference is given by
\be\label{massdiff:b}
\D m_{B_q}= 2|M_{12}^q|,\quad q= d, s\ ,
\ee
which is valid for $\G_{12}^q/M_{12}^q\ll 1$. Recall from \sec{general}
that, in order for a model with a real 
CKM matrix to be consistent with present experimental data,   
a new physics contribution to $\D m_{B_d}$ equalling at least 
$20\%$ of the SM contribution is necessary when $\r>0$. 
It can be shown that such a new contribution due to 
$M_{12}^{\mathrm{New}}|_d$ can be obtained in our model, 
while satisfying the bound of \eq{mass:diffK}. In fact, using 
Eqs.~(\ref{m12:bd}), (\ref{replacement}), (\ref{m12:Knew}), 
(\ref{m12:gen}), and (\ref{massdiff:b}), 
as well as central values for the various input 
parameters, one finds that even $(\b_d/\b_K)^2\sim \O(1)$ yields 
the necessary new physics contribution to $\D m_{B_d}$.
Moreover, it is clear from the preceding discussion that there is a 
relation between $\theta_d$ and $\theta_K$ in the above 2HDM. 
Indeed, employing \eqs{m12:Knew}{formula:epsi}, and (\ref{massdiff:b}), 
we obtain
\be\label{Bsys:sintheta}
\sin 2\theta_d = 2\sqrt{2}|\epsi_K|\frac{\D m_K}{\D m_{B_d}}
\left(\frac{m_{B_d}}{m_K}\right)^3\left(\frac{f_{B_d}}{f_K}\right)^2
\left(\frac{\b_d}{\b_K}\right)^2\ ,
\ee
and taking $f_K= 160\ \GeV$, $f_{B_d}=200\ \MeV$, we estimate 
\be\label{beta:bounds}
\left(\frac{\b_d}{\b_K}\right)^2\leqslant 7.4\ .
\ee
We see that, due to the enhancement factor of $(m_{B_d}/m_K)^3$ in 
\eq{Bsys:sintheta},  $\cp$ violation in $B_d^0$-meson decays 
is not necessarily small.
Before proceeding to discuss the $\cp$ asymmetries, we should  
mention that a more complete analysis of  $\cp$ violation and  
physics beyond the SM has to take into account the recent measurement of 
$\epsi'/\epsi_K$ \cite{KTeV}.  
However, as pointed out by Buchalla \ea\ \cite{buchallaetal} 
in the context of a specific 2HDM, a thorough investigation of the 
various contributions to direct $\cp$ violation in kaon decays, 
including renormalization group effects, is more involved  
and beyond the scope of the present paper.\bem{See also the discussion in 
\rf{luis}.}

\subsection{CP asymmetries and the 2HDM}
We now turn to a study of the $\cp$ asymmetries. Due to the fact that the 
new physics contribution has a complex phase, the $\cp$ asymmetries in 
different $B$-meson decay channels are also affected. For example,  
using \eq{Bsys:sintheta} and  the experimental value of \eq{exp:epsi}, 
we obtain for the $B_d^0$ system
\bea\label{thetasubd}
\sin 2\theta_d \simeq  0.13 \left(\frac{\b_d}{\b_K}\right)^2
\left(\frac{f_{B_d}}{200\ \MeV}\right)^2
\left(\frac{160\ \MeV}{f_K}\right)^2\ .
\eea
Recall from \eq{asym} that in the case of a real CKM matrix, 
with $\a=\p$ and $\b=\g=0$, the $\cp$ asymmetries take the form
\be
a_{J/\psi K^0_S}= \sin 2(\theta_d+ \theta_K), \quad 
a_{\p^+\p^-} = \sin 2(\pi -\theta_d)\ ,
\ee
and assuming the limit $\theta_K\ll \theta_d$, one finds
\be\label{predcn:multi}
a_{J/\psi K^0_S}= - a_{\p^+\p^-}\ .
\ee
We should emphasize that this result is a special feature of 
multi-Higgs-doublet models in which the CKM phase is absent,
provided the new physics contributions cannot 
compete with the $W$-mediated tree diagrams of the SM.
Note that due to the arbitrariness of the ratio $\b_d/\b_K$ in \eq{thetasubd},
 the $\cp$
asymmetries in this model can in principle vary from zero to one. 
This should be compared with the 
SM estimate which predicts $a_{J/\psi K^0_S}$ to lie within the 
range $0.5 \leqslant a_{J/\psi K^0_S}\leqslant 0.9$ \cite{parodi:etal,SM:UT}.
Once the two asymmetries $a_{J/\psi K^0_S}$ and $a_{\p^+\p^-}$ are measured, 
\eq{predcn:multi} will provide a clear test of the class of models where $\cp$ violation arises exclusively from flavour-changing neutral Higgs exchange.

\section{Discussion and conclusions}\label{summary}
One of the fundamental open questions in particle physics concerns the  
origin of $\cp$ violation. In particular, it is crucial to verify 
whether charged weak interactions violate $\cp$ or not.

We have analysed the viability of a real CKM matrix, considering 
our present knowledge of the CKM matrix, derived from strange and 
$B$-meson decays, the experimental value of $\D m_{B_d}$, and  
the improved bound on $\D m_{B_s}$. We have shown that if one assumes 
that physics beyond the SM contributes only to $\epsi_K$, then 
the real CKM matrix is clearly disfavoured by the above-mentioned data. 
However, if one makes the more 
plausible assumption that new physics also contributes to $\D m_{B_d}$ 
(and $\D m_{B_s}$), a real CKM matrix is still in keeping with present data. 
As a matter of fact, in order to fit the currently available data,
it is crucial to have a new contribution to $\D m_{B_d}$ corresponding to 
at least $20\%$ (for $\r > 0$),  while a new contribution to 
$\D m_{B_s}$ is not relevant at this stage. This is because only an 
experimental lower  bound on the oscillation frequency in the $B_s^0$ 
system is available.

We have presented a general analysis which is quite model 
independent and applicable to a large class of models.
This includes the supersymmetric SM with spontaneous $\cp$ violation, which 
has a real CKM matrix provided that only two standard Higgs doublets are 
introduced, 
apart from an arbitrary number of Higgs singlets. 
It is worth mentioning that within the supersymmetric SM it is 
possible to have the required additional contribution to 
$\D m_{B_d}$ \cite{gustavo:etal}.

Moreover, we have shown that the assumption of spontaneous $\cp$ violation 
plays an extremely important role when the naturalness of a real CKM matrix
is studied.  For illustration, we have investigated a specific 
two-Higgs-doublet model  where $\cp$ is spontaneously 
broken and the CKM matrix is real as a result of a $Z_2$ symmetry. 
In fact,  we have shown that one can generate $\epsi_K$ with simultaneous 
contributions to $\D m_{B_{d,s}}$ and obtain a good fit to the data.

We are eagerly awaiting the upcoming experiments at the various 
$B$ factory facilities which will provide further constraints on the 
standard mechanism of $\cp$ violation as well as on alternative scenarios of
$\cp$ breaking.  

\begin{ack}
The work of F.\,K. has been supported by the TMR Network of the EC under contract ERBFMRX-CT96-0090.
\end{ack}
%
%%%%%%%%%%%%% BIBLIOGRAPHY %%%%%%%%%%%%%%%%%%%%%%%%%%
%

%
%%%%%%%%%%%%%%%%%%%%%%%%%%%%%%%%%%%%%%%%%%%%%%%%%%%
%
%		FIGURE CAPTIONS
%
%%%%%%%%%%%%%%%%%%%%%%%%%%%%%%%%%%%%%%%%%%%%%%%%%%%
%
%\newpage
%\centerline{\bf FIGURE CAPTIONS}
%\begin{enumerate}
%\item[\bf Figure 1]The allowed region (at $\cl{95}$) in the $(\r, r_d^2)$ 
%plane for the Wolfenstein parameter $\r>0$ (a) and $\r<0$ (b), using 
%the experimental values for $|V_{ub}/V_{cb}|$, $|V_{cb}|$,  and $\D m_{B_d}$. 
%Each contour corresponds to a variation of the theoretical input 
%parameters, as described in the text [cf.~\eq{theo:var1}].
%\end{enumerate}
%
%%%%%%%%%%%%%%% FIGURE 1 %%%%%%%%%%%%%%%%%%%%%%%%%
%
%\newpage
%\begin{figure}
%\centerline{\hspace{-.4cm}\psfig{figure=f1.eps,height=3.7in,angle=0}}
%\vspace{0.9cm}
%\centerline{\psfig{figure=f2.eps,height=3.7in,angle=0}}
%\caption[]{\label{fig1}}
%\end{figure}
%
%%%%%%%%%%%%%%%%%%%%%%%%%%%%%%%%%%%%%%%%%%%%%%%%%%%
%
\end{document}